\documentclass[onecolumn,12point]{article}
\usepackage{amssymb}
\usepackage{graphicx}
\usepackage{enumerate}
\begin{document}

\title{Uncertainty Principle and the Standard Quantum Limits \thanks{This paper is published in Japanese translation in the October 2005 issue
of the Japanese magazine \emph{Mathematical Sciences}, pp. 35-40. It is put up on the quant-ph archive because it may be of interest to a wider audience.}}
\author{Horace P. Yuen\thanks{Email: yuen@ece.northwestern.edu} \\ Center for Photonic Communication and Computing\\
Department of Electrical and Computer Engineering\\
Department of Physics and Astronomy\\
Northwestern University, Evanston, IL 60208}

\maketitle

\begin{abstract}

The role of the Uncertainty Principle is examined through the
examples of squeezing, information capacity, and position
monitoring.  It is suggested that more attention should be
directed to conceptual considerations in quantum information
science and technology.
\end{abstract}

\section{Introduction}
In this article, I will outline my direct involvement with the
Uncertainty Principle through my research work on squeezed states
and the Standard Quantum Limit for monitoring the position of a
free mass.  More broadly, the Uncertainty Principle is connected
with general quantum limits on the information one can extract
from an otherwise noiseless classical system. This connection will
be highlighted with the classical capacity problem of a free
bosonic channel.  Some general comments on the emerging quantum
information science will be included.  There are a lot more things
I would like to discuss, but they have to be postponed to some other
publication due to space-time limitation.

\section{Squeezed States and Uncertainty Relations}
It is interesting to note that there is a continuing series of
international conferences under the name ``Squeezed States and
Uncertainty Relations.''  But I myself investigated squeezed
states through the study of quantum communication theory [1] for
classical information transmission, not through consideration of
the uncertainty relation although that certainly provided some
intuitive motivation.  The reason is that the usual mathematical
expression of the Uncertainty Principle, i.e., the Kennard
uncertainty relation
\begin{equation} \label{ur}
\langle\Delta a_1^2\rangle \langle\Delta a_2^2\rangle \geq 1/16,
\end{equation}
or its Robertson generalization rarely provides by itself the
solution of any problem.  (In (1), we have used the notation
$a=a_1+\iota a_2, a_1=a_1^{\dag}, a_2=a_2^{\dag}$, for the
destruction operator $a$ of a single boson mode.) This is because
each problem has its own figure of merit or criterion of success,
which is rarely a simple function of the second moments in (1).

For example, in optical communications which I considered, the
laser sources are usually taken to produce coherent states
$|\alpha\rangle$ in the ideal limit, with slowly fluctuating phase
and amplitude even in practice.  Thus, the following Standard
Quantum Limit (SQL) for coherent states applies,
\begin{equation} \label{sql}
\langle\Delta a_1^2\rangle_{SQL} \langle\Delta a_2^2\rangle_{SQL}
\geq 1/4. \end{equation} Clearly, the uncertainty relation (1)
allows the SQL (2) to be broken.  Indeed, one may have
$\langle\Delta a_1^2\rangle \rightarrow 0$ with corresponding
$\langle\Delta a_2^2\rangle \rightarrow \infty$, and with (1)
satisfied with equality for the so-called minimum uncertainty wave
packets, a subclass of pure squeezed states or what I called
two-photon coherent states (TCS). However, under a typical
constraint that only states $\rho$ of a maximum energy are
permitted, $tr \rho a^\dag a \leq S$, it is not a priori clear
that squeezed states would do better than coherent states for any
usual performance criterion.  This is because the high-noise
quadrature $a_2$ would consume a lot of otherwise useful mean
energy for the low-noise quadrature $a_1$, in order that $a_1$ be
squeezed, i.e., breaking the $\langle\Delta a_1^2\rangle_{SQL}
\geq 1/4$.

It turns out that squeezed states are indeed better [1] if the
total energy is distributed properly as mean and fluctuation ones.
For example, they improve the coherent-state single quadrature
signal-to-noise from $4S$ to $4S(S+1)$.  This means that for the
estimation of the mean quadrature with homodyne detection, the
\emph{rms} error is improved from $1/\sqrt{S}$  to $1/S$. Please
see Ref [1] and references cited therein for a fuller treatment.

Unfortunately, the impact of squeezed states is severely limited
by the inevitable loss involved in a real system.  Upon a linear
loss of $1-\eta$ so that $\eta$ is the transmittance, any
squeezing $\langle\Delta a_1^2\rangle$ becomes
\begin{equation}
\langle\Delta a_1^2\rangle \rightarrow \eta \langle\Delta
a_1^2\rangle + (1-\eta)/4,
\end{equation}
 which is effectively wiped out in large
loss.  Even when $1-\eta$  is small, large squeezing still becomes
impossible.  This \emph{sensitivity} issue could be even worse in
the newer area of quantum information science and technology,
whenever quantum entanglement plays an essential role.  Thus, in
quantum computation with multiparticle entanglement, small loss
would effectively destroy the entanglement essential for the
computation, similar to what I would call the
\emph{supersensitivity} of macroscopic superposition of quantum
states to loss.  This means a multi-qubit superposition would
effectively decohere when just one qubit state moves out of the
two-dimensional space of its original description to an orthogonal
direct summand of a larger state space including loss. This issue
has not, to my knowledge, been properly treated theoretically in
the literature.  In particular, quantum leak plumbing is not
sufficient.  I believe loss is a formidable real obstacle to a
realistic implementation of quantum computation via multiparticle
entanglement, because it cannot be corrected, or at least not yet
shown to be correctable, by utilizing more qubits as in the usual
model of fault-tolerant quantum computing.  A detailed analysis
will be presented elsewhere.

\section{Bosonic Channel Capacity}

The classical information transmission capacity of a free bosonic
channel under a quantum state energy constraint can be derived
through the entropy bound, often called the Holevo inequality, the
most general version of which was first derived by Ozawa [2]
through Lindblad's inequality that has since become the most
powerful approach in the subject.  Under $tr \rho a^\dag a \leq S$
, the capacity is
\begin{equation} \label{capacity}
C(S)=(S+1)\log(S+1) - S\log(S).
\end{equation}
This result has since been generalized to a linear lossy channel
[3], but still resists further generalization, e.g., to include an
additive classical noise.

This capacity result is usually viewed as the consequence of
quantization of an otherwise continuous classical field mode.
Thus, the capacity is finite for a finite energy due to the
quantization of energy levels for $N = a^\dag a$, in contrast to
infinity in the ideal classical limit. But is there any role for
the Uncertainty Principle in this kind of result? The answer is
yes [2].  If the energy Hamiltonian is $H=P^2$, e.g., instead of
$N$, where the $P \sim a_2$  could be the momentum of a free
particle, boson or fermion, the capacity is unlimited as in the
classical situation under the constraint $tr \rho H \leq S$ .
However, the uncertainty relation (1) puts a limit on the
realizable capacity when the spatial extent of the system, or
equivalently $\Delta Q$, is limited. This built-in ``finitism'' of
the quantum case is very satisfactory from an intuitive physical
point of view, and is enforced by uncertainty relation of the form
(1) for conjugate observables that have continuous spectra. While
there is no quantization per se for one observable, there is
nevertheless a finite limit if no other physical infinity is
allowed.  This is similar to the constraint of (1) on producing
squeezing - useful energy needs to be spent on the other
quadrature.

\section{Standard Quantum Limit for Position Monitoring}

As can be seen from the other papers of this special issue, a main
problem for the physical interpretation and application of the
Uncertainty Principle
\begin{equation}
\Delta Q \Delta P \geq \hbar
\end{equation}
concerns the possible meaning of $\Delta Q$  or $\Delta P$  as the
result of disturbance on the system produced by a measurement. If
we interpret $\Delta Q$  and  $\Delta P$  rather formally as
standard deviations, the above inequality was shown by Kennard to
be universally valid, and this interpretation is often referred to
as the uncertainty relation.  On the other hand, I believe the
interpretation of $\Delta Q$ and $\Delta P$ as measurement noise
and disturbance lies behind the development of the Standard
Quantum Limit (SQL) for monitoring the position of a free mass,
which is applicable to gravitational wave detection. The SQL
states that [4,5] if the position of a free mass $m$ is measured
at $t=0$, the position fluctuation at $t>0$ is at least:
\begin{equation}
\langle \Delta X^2(t) \rangle_{SQL} = {\hbar t}/m.
\end{equation}
 The
derivation of (6) was taken to be universally valid as a
consequence of the Uncertainty Principle, and it was concluded
that the free mass position is not a ``QND observable'' - namely,
that the position measurement cannot be a ``quantum nondemolition
measurement'' because the disturbance to the system from the first
position measurement demolishes the possibility of an accurate
second measurement after an interval t of free evolution.  It was
pointed out [6] that the derivation of (6) from the Uncertainty
Principle is incorrect, and in fact (6) needs not hold at all.  In
the following, a brief qualitative discussion will be provided.  A
detailed quantitative description is given in [1], and full treatment in
various papers of Ozawa referred therein.

The usual textbook description of quantum measurement is grossly
incomplete.  In the first place, the measurement probability is
not just generated by a selfadjoint operator or equivalently a
projection-valued measure (PVM) on the system state space, but
rather by the more general positive operated-valued measure (POM).
More significantly in this context, the state after measurement
needs not be the same as the one whose projection gives the
measurement probability.  In the nondegenerate case, one may
describe this by the ``dyad''
\begin{equation}
|\Psi_s\rangle\langle\Psi_m|
\end{equation}
to describe a measurement result $m$ from measurement on a system
in state $|\Psi\rangle$, with $|\langle\Psi_m|\Psi\rangle|^2$ the
probability of obtaining $m$ and $|\Psi_s\rangle$ the system state
immediately after measurement, which depends on $m$ in general.
When $|\Psi_s\rangle=|\Psi_m\rangle$, the measurement is called
``the first kind'' by Pauli [7], and is often called  ``quantum
nondemolition measurement'' nowadays, adding confusion to the QND
terminology [4,5] above.  Pauli calls a measurement the ``second
kind'' if it is not of the first kind. The general measurement
description is given by Ozawa [8] in his concept of a ``completely
positive instrument'', by which he proved, as a special case of a
more general result, that every ``dyad'' could be realized in
principle via an interaction between the object and a measuring
apparatus.

It is easy to see that if
$|\Psi_s\rangle=|\Psi_m\rangle=|x\rangle$, the position eigenstate
(going outside the Hilbert Space framework with Dirac notation), a
precise position measurement leaves the system with infinite
momentum fluctuation, and so no accurate second position
measurement is possible.  Even for nonexact position measurement,
(5) is supposed to yield a correspondingly large momentum
disturbance to hinder the next position measurement.  The actual
derivation of (6) did not utilize such interpretation, which was
just taken to be the intuitive reason for its validity, but was
rather deduced as a mathematical consequence of the uncertainty
relation obtained by interpreting the disturbance $\Delta Q$ and
$\Delta P$ in (5) as standard deviations, and that is of
course incorrect. It is clear from (7) that measurement of the
second kind would not suffer from such disturbance, and thus the
SQL cannot be derived from the uncertainty relation, which is
merely a generally valid relation on a quantum state.  I was
indeed led by such consideration to the rejection of the SQL as a
universal law.

The general problem of disturbance and noise is multi-faceted, as
there are different definitions of ``disturbance'' as well as
``noise'' that are appropriate under different situations.  A long
way toward the clarification and elaboration of these problems has
been covered by the work of Ozawa, as described elsewhere in this
issue and in the papers referred to therein.  It is fair to say at
this point that there is no longer any excuse to confuse intrinsic
state fluctuation and action-induced disturbance in quantum
physics.

\section{Perspective}
Noise and disturbance play an essential role in the security of
the BB84 type quantum cryptographic protocols.  The noise an
eavesdropper suffers is inverse monotonically related to the
disturbance she introduces, which can be measured by the users.
Security is obtained when the users are assured that the
disturbance is below a threshold that would allow them to
eliminate her information obtained from her noisy measurement.
The problem does not directly fit the noise-disturbance relations
obtained so far, but Ozawa is making progress in this direction
using his inequality.

I would like to suggest that in physical and engineering sciences,
there are three kinds of considerations that one may entertain
that are quite distinct:
\begin{enumerate}[(i)]
\item mathematical \item    physical \item   conceptual.
\end{enumerate}

The first refers to precise mathematical relations of the kind
current in modern mathematics, and not to symbolic calculations.
While the mathematical abstraction often seems ``unphysical'', the
focus on essentials and general possibilities could be very
powerful tools for solving concrete problems.  The second refers
to the usual intuition a physicist or engineer develops on his
subject, of the kind that directly involves the concrete entities
of the subject.  I think the distinction between the first two
kinds is rather clear, although the relation between mathematics
and the physical world is intricate and forever fascinating.  Some
discussion on the role of mathematical rigor in actual
applications to the world is given in  [9].

The third refers to a kind of thinking on concepts that are
neither mathematical nor physical, but which tie together the two.
Examples include concepts from communication theory, information
theory, and cryptography, in particular as they relate to the
working of real systems.  Of special importance is how a problem
from the real world, which is not already formulated
mathematically and which in fact does not allow an
all-encompassing mathematical formulation, can be conceptualized
to allow mathematical and physical representations suitable for
different purpose.  The purpose dictates whether all the essential
features are included in the representation, so that conclusions
drawn from it are indeed relevant to the purpose according to its
success criteria.  As a specific example, what is the operational
significance of the entropy of a bit string in the context of
privacy from an attacker?  In the context of communication, it has
been related to the operational or empirical quantities of error
rate and data rate.  But what about in cryptography?  See [10] for
an illustration on the inadequacy of entropy as a quantitative
measure of security.  The key bit-string is not necessarily secure
even if the attacker's information about it is small but not
exponentially small in the bit-string length, which in turn cannot
be achieved in general by privacy amplification.  A detailed
demonstration is forthcoming.

Quantum information science and technology has the distinction
that all these three kinds of consideration are crucial in many
problems, in contrast to more traditional areas.  Careful
conceptual thinking, as distinct from mathematical or even
physical intuition, is already essential in fundamental
considerations of many subjects including physics.  One may say, I
believe, Heisenberg was confused in his original elaboration of
the Uncertainty Principle about the fluctuation of the system
before measurement and after measurement.  His physical intuition
from his well-known microscope example was not conceptually
sharpened, and is not properly expressed mathematically either by
his original expression or by the Kennard inequality.  His
confusion has influenced generations of physicists, including and
at least up to the SQL.

It seems to me that current quantum information science and
technology also suffers from a number of inadequacies in its
foundation.  In particular, many mathematical models that have
been extensively analyzed are not sufficiently connected to
physical and conceptual considerations on realistic experimental
situations that would allow one to draw useful conclusions for
real applications.  In this paper, I have indicated the examples
of loss in quantum computation and of entropy in cryptography.
Other examples abound, of which I may mention quantum bit
commitment [11] for which the characterization of a bit commitment
protocol has not been clarified in the impossibility claim.  While
this claim may shut off a useful area prematurely, the other
deficiencies may divert resources and effort in the wrong
direction.  It would be good for both intellectual and practical
reasons to have more attention directed to conceptual
considerations, especially ones related to modeling and
sensitivity.  It is important to remember that physicists and
engineers need quantitative theories, not just qualitative or
asymptotic ones, to build real systems.  In particular, the
practical requirement of robustness to small imperfections has to
be thoroughly investigated for quantum-entangled systems.

Miyamoto Musashi tells the readers of his ``The Book of Five
Rings'' to consider the principles presented as though they were
discovered from their own minds.  In the case of studying a
scientific subject, this may perhaps be interpreted as urging one
to think through the foundation of the subject and the logical
interconnection of the principles in a way the creator of the
principles may have gone through.  In particular, one could have
actually discovered some of these principles himself from such
consideration before learning them.  This kind of process is
essential for true understanding from my own experience, and no
doubt that of many others.  I believe it is of especial importance
in quantum information science and technology.

\section{Acknowledgements}

I would like to thank Professor Ozawa for many discussions on
these topics, and for kindly translating this paper into Japanese.


\begin{thebibliography}{10}
\expandafter\ifx\csname url\endcsname\relax
  \def\url#1{\texttt{#1}}\fi
\expandafter\ifx\csname
urlprefix\endcsname\relax\def\urlprefix{URL }\fi

\bibitem{yuen03} H. P. Yuen,``Communication and Measurement with Squeezed
States'', in \emph{Quantum Squeezing}, edited by P.D. Drummond and
Z. Ficek, Springer Verlag 2003, pp. 227-261.

\bibitem{yuenozawa} H. P. Yuen and M. Ozawa, Phys. Rev. Lett. 70, 262 (1993).

\bibitem{yuenetal} V. Giovannetti, S. Guha, S. Lloyd, L. Maccone, J. H. Shapiro,
and H. P. Yuen, Phys. Rev. Lett., 92, 027902 (2004).

\bibitem{brag} V. B. Braginskii and Yu. L. Vorontsov, Sov. Phys. Usp. 17, 644
(1975).

\bibitem{caves} C. M. Caves, et al., Rev. Mod. Phys. 52, 341 (1980).

\bibitem{yuen83} H. P. Yuen, Phys. Rev. Lett. 51, 719 (1983).

\bibitem{pauli} W. Pauli, Handbuch der Physik, vol 5, Springer, 1958.

\bibitem{ozawa} M. Ozawa, Journal of Math. Phys. 25, 79 (1984).

\bibitem{yuen97} H. P. Yuen, in \emph{Quantum Communication, Computing and
Measurement}, ed. by O. Hirota, etc., Plenum, New York, pp. 17-23,
1997.

\bibitem{yuenqph}
 H. P. Yuen, ``KCQ:  A New Approach to Quantum Cryptography I.
General Principles and Qumode Key Generation'', quant-ph 0311061.

\bibitem{qbc}
H. P. Yuen, ``How to Build Unconditionally Secure Quantum Bit
Commitment Protocols'', quant-ph 0305144.



\end{thebibliography}
\end{document}